# Direct observation of the layer-number-dependent electronic structure in few-layer WTe$_2$

Masato Sakano[1*], Yuma Tanaka[1*], Satoru. Masubuchi[2*], Shota Okazaki[3], Takuya Nomoto[1], Atsushi Oshima[1], Kenji Watanabe[4], Takashi Taniguchi[5], Ryotaro Arita[1,6], Takao Sasagawa[3], Tomoki Machida[2], and Kyoko Ishizaka[1,6]

[1]Quantum-Phase Electronics Center and Department of Applied Physics, The University of Tokyo, Bunkyo-ku, Tokyo, 113-8656, Japan
[2]Institute of Industrial Science, The University of Tokyo, Meguro-ku, Tokyo 153-8505, Japan
[3]Materials and Structures Laboratory, Tokyo Institute of Technology, Yokohama, Kanagawa, 226-8503, Japan.
[4] Research Center for Functional Materials, National Institute for Materials Science, 1-1 Namiki, Tsukuba 305-0044, Japan
[5]International Center for Materials Nanoarchitectonics, National Institute for Materials Science, 1-1 Namiki, Tsukuba 305-0044, Japan
[6]RIKEN Center for Emergent Matter Science (CEMS), Wako, Saitama, 351-0198, Japan
*These authors equally contributed to this work

**When a crystal becomes thinner and thinner to the atomic level, peculiar phenomena discretely depending on its layer-numbers ($n$) start to appear. The symmetry and wave functions strongly reflect the layer-numbers and stacking order, which brings us a potential of realizing new properties and functions that are unexpected in either bulk or simple monolayer. Multilayer WTe$_2$ is one such example exhibiting unique ferroelectricity and non-linear transport properties related to the antiphase stacking and Berry-curvature dipole. Here we investigate the electronic band dispersions of multilayer WTe$_2$ (2–5 layers), by performing laser-based micro-focused angle-resolved photoelectron spectroscopy on exfoliated-flakes that are strictly sorted by $n$ and encapsulated by graphene. We clearly observed the insulator-semimetal transition occurring between 2 and 3-layers, as well as the 30-70 meV spin-splitting of valence bands manifesting in even $n$ as a signature of stronger structural asymmetry. Our result fully demonstrates the possibility of the large energy-scale band and spin manipulation through the finite $n$ stacking procedure.**

The physical property in the atomically-thin two-dimensional (2D) material drastically changes as the number of layers decreases to the monolayer limit[1]. The quantization of band dispersions along the stacking direction, as well as the reduction of symmetry compared to the infinite bulk crystal can strongly modify the electronic structures of 2D materials, resulting in the peculiar physical phenomena such as the sequential band gap variation in phosphorene[2,3] and the spin–valley coupling in $MoS_2$[4-8]. $WTe_2$ (Fig. 1a) is also a striking material f[9-12]. Bulk $WTe_2$ is a charge-compensated polar semimetal exhibiting non-saturating magnetoresistance[13] and had been predicted as an inversion-symmetry-broken type-II Weyl semimetal[14-16]. In contrast, mono- and 2-layer $WTe_2$ are known as insulators with small bandgaps[9-11,17,18]. Particularly the monolayer $WTe_2$ is known as the 2D quantum spin Hall insulator[9,11,17,18]. In terms of the structural symmetry, although the monolayer $WTe_2$ is centrosymmetric, the stacking order of layers breaks the inversion-symmetry in few-layer $WTe_2$[10,11,19-22], leading to the unique phenomena such as the nonlinear anomalous Hall effect (NAHE)[12,20,23,24] and the ferroelectric switching[10,12] related to Berry-curvature dipole. Furthermore, among the few-layer $WTe_2$, the even-odd layer-number dependence in the NAHE through the ferroelectric switching was theoretically predicted[22], and recently experimentally demonstrated in the 3- and 4-layer $WTe_2$[12].

The abovementioned studies on thin-flake $WTe_2$ have been mostly proceeded by comparing the transport properties on micro-devices with the band calculations. Due to the limited size (typically ~10 μm) and volume of the samples as well as the difficulty in handling, many of the physical property measurements established in bulk systems are not available. Nevertheless, the precise investigation of the electronic structure is extremely important, especially considering that the property related to Berry-curvature is determined by the electronic band dispersions in the reciprocal space. We should note that band calculations are recently making great progress, and a variety of methods including Heyd–Scuseria–Ernzerhof (HSE) hybrid functional[25] which is known to reasonably reproduce the band gap size are now frequently used for atomically thin flakes. Still there are many difficulties specific to flakes; the lack of structural information (strict atomic positions), the computational cost of handling the finite-size slabs, and so on. Thus the experimental clarification of the band structures is highly desired. Recently, a micro-focused angle-resolved photoemission spectroscopy (μ-ARPES) study by Cucchi *et al.*[11] successfully captured the electronic structures of the mono- and 2-layer $WTe_2$ by scanning the stairstep-like exfoliated $WTe_2$ flake encapsulated by graphene. In this work, we developed an efficient procedure to prepare the series of graphene-encapsulated $WTe_2$ multilayer flakes ideally sorted by layer-numbers, by utilizing 2D materials manufacturing system (2DMMS)[26,27]. By further using the laser-based μ-ARPES system (see Method), we unambiguously clarified the electronic band structures in 2–5 layer $WTe_2$. It clearly revealed the insulator–semimetal transition realized between the 2- and 3-layer, and the striking even–odd effect in the layer-number dependence of the spin-band splitting.

**Sample fabrication and characterization**

The samples were fabricated by an all-dry pick up and flip method using Elvacite2552C copolymer inside a glovebox chamber[26,27] as shown in Fig. 1b-d (see Methods for details). Hexagonal boron nitride (hBN), graphite, WTe2, and graphene flakes were sequentially picked-up by a polymer stamp (Fig. 1b). The assembled heterostructure was once transferred to another polymer stamp at room temperature to turn over the heterostructure face (Fig. 1c) and dropped onto a SiO2/Si substrate with a prepatterned metal electrode (Fig. 1d). Figure 1e shows an optical microscope image of a typical device consisting of WTe2 flake covered by a few-layer graphene. The encapsulation by the graphene flake provides a sufficiently clean surface for the ARPES measurements[11,28]. The number of layers of the fabricated samples was first distinguished by analyzing the optical microscope images based on deep learning [27,29], which was further unambiguously determined by the obtained ARPES spectra. The upper panels in Fig. 1f show the 2–5 layer WTe2 flakes (shown with white frames) used to obtain the corresponding ARPES images in the lower panels, respectively. As the thickness of the flake increases, the number of the bands clearly increases. Figure 1g shows the energy distribution curves (EDCs) at $k_x = 0$ for 2–5 layer WTe2, obtained from the ARPES images shown in Fig. 1f. From the number of the intensity peaks in the energy region $E - E_F = -0.15$ to $-0.45$ eV, we can determine $n$ as it well corresponds with the quantized eigenstates reflecting the number of the stacked layers. From this we can confirm that the 2–5 layer WTe2 flake samples are successfully prepared and separately measured by µ-ARPES.

**Insulator-semimetal transition between 2- and 3-layer WTe2**

First, we focus on the electronic structures near the Fermi level ($E_F$). Figures 2a–d show the ARPES images for 2–5 layer WTe2, divided by the Fermi–Dirac distribution function convolved with the Gaussian function of 3 meV, to remove the effect of intensity cutoff near $E_F$. Here the $k_x$ axis corresponds to the Γ-X direction, where the semimetallic characters with hole and electron bands are most well observed in bulk WTe2[13-16,30,31]. In these figures, we can clearly distinguish the $n$-dependent evolution of the hole-like band dispersions around $k_x \simeq -0.2$ Å$^{-1}$, seemingly varying from being fully occupied ($n = 2$), nearly touching $E_F$ ($n = 3$), to apparently crossing $E_F$ ($n = 4$ and 5).

In 2-layer WTe2, the valence band maximum is actually located at the Γ point where the intensity is weak. By closely looking at the EDCs, its energy position is estimated to be −22(5) meV (see Supplementary section S1). We note that the hole-like flat band feature around $k_x \simeq -0.2$ Å$^{-1}$ and −35 meV is in a good agreement with the previous ARPES result[11], while the electron band previously observed around −0.3 Å$^{-1}$ is absent in our result recorded at 27(8) K. For comparison, we show in Fig. 2e the result of band calculation utilizing the HSE[25] hybrid functional method (red curves) overlaid with that of the generalized gradient approximation method proposed by Perdew, Burke and Ernzerhof (GGA-PBE)[32] (dashed blue curves) (see Methods for the details). Although the hole-like bands around −0.2 Å$^{-1}$ are more dispersive than those experimentally obtained, the formation of the small band-gap (~40 meV) between the electron-like (above $E_F$) and the hole-like band (below $E_F$) is well reproduced by the HSE calculation, in contrast to GGA-PBE. Thus our result indicates that $n = 2$ WTe2 flake belongs to the gapped insulator phase with fully occupied hole bands. To experimentally detect the possible electron-like band in the unoccupied side, we further performed ARPES measurement at room temperature (Fig. 2f). Dividing by the 300 K Fermi–Dirac distribution function allows us to visualize the electronic structures in $E \leq E_F + 100$ meV. Figure 2g shows the EDCs extracted from the ARPES image in Fig. 2f, with respective peak positions plotted by markers. In $k_x = 0 \sim -0.2$ Å$^{-1}$ region, the dispersion of the valence band edge appears around 30-60 meV below $E_F$, whereas the contribution of the electron band bottom shows up around $k_x = -0.2$ to $-0.27$ Å$^{-1}$ region at ~50 meV above $E_F$. With this observation, in combination with the HSE calculation, we conclude that $n = 2$ WTe$_2$ has the gapped semiconducting band structure, consistent with the previous transport measurement[9].

To similarly discuss the detailed near-$E_F$ electronic structure in 3-layer WTe$_2$, we show in Fig. 2h,i the band calculation (HSE and GGA-PBE) and the ARPES image combining the data taken by the $p$- ($k_x < -0.25$ Å$^{-1}$) and $s$-polarized ($k_x > -0.25$ Å$^{-1}$) lights, respectively. We note that the electron band appears stronger with $p$-polarization setup because of the matrix-element effect. Figure 2j shows the EDCs extracted from the ARPES image in Fig. 2i, with the markers indicating the peak positions. At $k_x = -0.32$ and $-0.36$ Å$^{-1}$, faint but finite ARPES intensities (○ in Fig. 2j) show up near $E_F$, as compared to $k_x > -0.28$ Å$^{-1}$ region. It corresponds to the bottom of the conduction band crossing $E_F$, as observed by using the $p$-polarized light (Fig. 2i). The HSE calculation for 3-layer WTe$_2$ (red curve in Fig. 2h) also well reproduces such small overlaps of the electron- and hole-bands at the Fermi level. Our ARPES results thus spectroscopically confirm that the HSE calculation better reproduces the near-$E_F$ electronic structure as compared to GGA, at least for $n = 2, 3$. It also shows that the insulator–semimetal transition, appearing as the overlap of the hole and electron bands on increasing the number of layers $n$, is realized between 2- and 3-layer, being consistent with the transport property[9].

**Layer-number-dependent band spin splitting in 2–5 layer WTe$_2$**

Now we present the layer-number-dependent spin splitting of valence bands in few-layer WTe$_2$ that arise from the noncentrosymmetric structure. Figure 3a–d show the ARPES images of 2–5 layer WTe$_2$ recorded along the $k_x$ direction. Here we focus on the band dispersions in the energy range of –0.15 to –0.5 eV. On the right side ($k_x \geq 0$ region), the peak positions of ARPES intensities estimated from the EDCs and the momentum distribution curves (MDCs) are plotted with markers ● and ♦, respectively. When the spin splitting is small, it is hard to figure out whether the spectral peak consists of single or double components. In such a case we place a marker at the peak position estimated by assuming a single component, together with a bar whose ends represent the peak positions estimated by assuming double components. The obtained band dispersions are then indexed from the lower energy side as #$i$ ($i = 1-n$). As we can see in Fig. 3a-d, the bands in 2- and 4-layer WTe$_2$ clearly exhibit the notable band splitting as moving away from $k_x = 0$. This band splitting can be regarded as a strong evidence of inversion-symmetry breaking which lifts the spin degeneracy through the spin‑orbit interaction. Nevertheless, 3- and 5-layer WTe$_2$ exhibit small or indetectable band splitting as shown in Fig. 3b,d, thus raising the clear even-odd effect in the layer-number dependence. Considering that the structures of these few-layer WTe$_2$ all belong to the same point group (C$_s$) regardless of $n$[19,22], such even-odd effect cannot be understood in a simple symmetry-based argument.

Here we compare the experimental result with the GGA-PBE slab calculation (see Methods) that can fully handle 2-5 layer WTe$_2$, to quantitatively discuss the $n$-dependent valence band splitting. Figures 3e−h show the calculated band dispersions along the $k_x$-direction, with the red-blue colored weights indicating the spin polarizations along $y$ ($S_y$, left-side) and $z$ ($S_z$, right-side). Note that the spin polarization along $x$ is not allowed on the $k_x$-axis due to the mirror ($yz$ is the mirror plane) and time reversal symmetries. As increasing $n$, the number of valence bands also correspondingly increases, similarly to the ARPES results in Fig. 3a-d. When we focus on the $S_y$ and $S_z$ components as well as the band splitting, there are certain quantitative differences between the even- and odd-$n$ cases. One is that the energy of the band splitting does not show any monotonous change as a function of $n$: The spin splitting obtained in 2- and 4-layer WTe$_2$ is apparently larger than those in 3- and 5-layer. Another difference is found in the spin component of the band dispersions showing the splitting. For example, when we look at bands #1,2 for 2−5 layer WTe$_2$, the main spin component is predominantly $S_z$ for 2- and 4-layer, while it rather appears in $S_y$ for 3- and 5-layer. Similar trends are also confirmed for many other bands #$i$. In general, the spin−orbit interaction works as an effective magnetic field that is perpendicular to the potential gradient and to the momentum of electrons[8,33]. It thus indicates that on the $k_x$-axis, $S_y$- and $S_z$-components are respectively induced by the local potential gradient along $z$ and $y$ in the crystal structure. The result of GGA-PBE calculations in Fig. 3e-h, which well catches the features obtained by ARPES, thus suggests that the characteristics of the potential gradient associated with the spatial asymmetry alternately changes as $n$ is increased, resulting in the even−odd effect in the number-of-layers. We note that the HSE calculation on 2- and 3-layer WTe$_2$ also shows the similar trend (see Supplementary section S2). It should be noted that both GGA-PBE and HSE calculations performed here use the WTe$_2$ slab-models with the atomic coordinates fixed to those in the bulk crystal[34], thus ruling out any influences related to atomic relaxations or reconstructions specific to flakes. In bulk crystals, the WTe$_2$ layers stack on each other alternatively with $\pi$-rotation along $z$, i.e. keeping $z$ direction as the $2_1$ screw axis (see Fig. 1a). Our result suggests that this peculiar antiphase stacking should be playing the dominant role on the even-odd effect of the spin splitting, through the structural asymmetry as reflected in $S_y$ and $S_z$.

**Discussion: origin of the even-odd $n$-dependence**

Let us discuss the mechanism of the even–odd $n$ dependence appearing in few-layer WTe$_2$. To more quantitatively analyze the $n$-dependent spin band splitting, we focus on the EDCs at $k_x = -0.1$ Å$^{-1}$ as shown in Fig. 4a. The markers, horizontal bars, and the band indices #$i$ are given in the similar manner as those in Fig. 3a-d. The two-peak structures for bands #$i$ can be separately observed for even $n$ samples, whereas they are hard to discern in odd $n$ cases. In Fig. 4b, the energies of the band splitting estimated from the separation of these peak positions in Fig. 4a are summarized (solid circles ●) with those similarly obtained by GGA-PBE calculations at $k_x = -0.1$ Å$^{-1}$ (open circles ○). The values of band splitting in 2- and 4-layer WTe$_2$ are mostly spreading in the range of 20–70 meV, whereas those in 3- and 5-layer are limited within the range of 0–30 meV except for the calculated band #1 in $n = 5$. This result shows the peculiar even-odd effect in $n$-dependent band splitting.

To discuss the possible mechanism from the structural aspect, the side and top views of the $WTe_2$ crystal structure are shown in Fig. 4c,d. As already mentioned, the adjacent layers are stacked with $\pi$ rotation around $z$, thereby keeping the $2_1$ screw axis along $z$ in bulk. This is unique to the $T_d$ and $T'$ phases in $WTe_2$ and $MoTe_2$, and in a strong contrast to the undistorted $CdI_2$ structure (so-called $1T$-type) where the trigonal layers stack in in-phase order. Such out-of-phase ordered stacking is rather reminiscent of $2H$-type, as seen for example in $2H$-$WSe_2$ with the triangular-prism $WSe_6$ coordination. There, the spin-valley coupled band splitting reflecting the $C_3$ symmetry appears in monolayer, whereas the spin-valley polarization cancels out and regains the global inversion symmetry when the even-number layers are stacked[8,35,36]. This thus causes the peculiar even-odd $n$-dependence in the spin-valley polarizations. Here we make a similar consideration for $WTe_2$. First, for simplicity, let us consider a virtual model in which the undistorted $CdI_2$-type monolayers of $WTe_2$ (point group $\bar{3}m$ with no net polarization) stack with alternative $\pi$ rotations, without any in-plane shift. The schematics of this virtual model viewed from $x$-direction are depicted in Fig. 4e,f for $n = 2$ and 3. The black block indicates the unit of $WTe_2$ single layer, and the orange arrow indicates the direction of the triangular network of the top and bottom Te layers. The black marker × indicates the $C_2$ axis within each undistorted $WTe_2$ monolayer unit. In this virtual model, there is naturally no breaking of the inversion symmetry for $n = 1$. However, looking at the $n = 2$ case in Fig. 4e, the global $C_2$ axis is apparently lost due to the out-of-phase ordered stacking. For $n = 3$, on the other hand, the global $C_2$ axis recovers again in the second layer. Considering that the $yz$-plane is a mirror plane, this $n$-dependent $C_2$ symmetry around $x$-axis can explain the even-odd effect of the spatial asymmetry causing the spin splitting: The potential gradient along $y$ derived from the Te triangular network can cause the large $S_z$ polarizations on $k_x$-axis, without cancelling out in the even-number-layer cases.

We note that, in the real few-layer $WTe_2$, the situation should be much more complicated. The real atomic positions significantly deviate from those in $CdI_2$, and there are also slight layer-dependent shifts of the "$C_2$ axes" along $y$-directions, leading to the lack of the inversion center for all $n \geq 2$. Such spatial asymmetry along $y$ is known to strongly couple to that along $z$ (i.e. ferroelectric polarization) *via* the alternative stacking[19,22]. In addition, in reality, the $n$-dependent atomic relaxation may occur and more complicated band spin splitting involving both $S_y$ and $S_z$ could be realized, as compared to our calculations in Fig. 3e-h. Although such possibility cannot be completely ruled out, our discussion based on the ARPES result with GGA calculations safely conclude that the observed even–odd $n$ dependence of the band splitting is mainly attributed to the in-plane potential gradient by the triangular network of Te, contributing to the stronger structural asymmetry appearing in even $n$. Our study suggests the possibility that few-layer stacking can exhibit peculiar physical property that is absent in monolayer and bulk materials, as represented by the strong layer-number-dependent spin and electronic structures in the present $WTe_2$ case. The precise electronic structures revealed by ARPES thus offers us the direct information which should be useful for comprehending the variety of physical properties in 2D systems.

**Summary**

In conclusion, by combining the robotic sample fabrication using vdW assembly and the laser μ-ARPES systems, we precisely investigated the layer-number dependent band structures in isolated few-layer WTe$_2$ micro-flakes. By clarifying the *k*-dependent near $E_F$ structure, we spectroscopically determined that the insulator–semimetal transition is realized between the 2- and 3-layer. Furthermore, we found the strong even–odd *n* dependence of the spin band splitting in ~70 meV scale, originating from the in-plane potential gradient of Te triangular network. It successfully demonstrates the finite *n* effect appearing in the anti-phase stacking system, and raises the possibility of the large energy-scale band and spin manipulation through the finite *n* stacking procedure. The sample fabrication and preparation process used in this study can be applied to a variety of composite systems including twisted and hetero-structured van der Waals materials, and will enable us the direct observation of the complicated electronic structure associated with the novel properties and fuctions[37,38].

**Methods**

**Crystal growth**

Single crystals of $WTe_2$ were grown by a self-flux method[39]. High purity grains of W (99.9999%, 0.1 g) and Te (99.99999%, 5.0 g) were sealed in a vacuum-sealed quartz tube, completely melted at 900 °C, and slowly cooled down to 600 °C at a rate of 1 °C /h. The Te-flux was removed by centrifugation at 600 °C using quartz wool as a filter. In order to remove the remaining Te-flux on the crystal surface by sublimation and to improve the crystal quality by annealing, the obtained crystals were heat-treated in a vacuum-sealed quartz tube (430 °C at the crystal position and 100 °C at the other end) for 50 hours.

**Exfoliation and Searching for hBN, graphene, graphite, and $WTe_2$ flakes**

1. We exfoliated graphene and hBN flakes onto an $SiO_2$/Si substrate that had been treated by a plasma cleaner (Harrick Plasma PDC-32G). Monolayer or bilayer graphene, graphite, and hBN flakes were searched using the automated optical microscope integrated with deep-learning-based image recognition algorithm[27,28] in a glovebox chamber.
2. We exfoliated $WTe_2$ onto an $SiO_2$/Si substrate after an $O_2$ plasma treatment at 90 W for 3 min. inside a glovebox chamber. The 2–5 layer $WTe_2$ flakes were searched using the automated optical microscope.
3. hBN, graphene, and graphite flakes were annealed in $Ar/H_2$ (97:3) gas at 600 °C for 3 hours to remove the contaminations from their surface.

**Assembly**

4. The hBN flake was picked-up by Elvacite2552C polymer on a glass slide at 80 °C. The pick-up process was repeated with graphite, $WTe_2$, and graphene flakes at 80, 90, and 80 °C, respectively.

**Flip**

5. Elvacite2552C and ionic liquid (Elvacite2552C + ion) were mixed and prepared on a glass slide. We found that adding a small portion of ionic liquid to the Elvacite2552C greatly enhances the adhesion to 2D flakes. Therefore the graphene/$WTe_2$/Graphite/hBN stack can be transferred to the Elvacite2552C + ion from the Elvacite2552C at room temperature. The heterostructure on Elvacite2552C + ion was taken out of glovebox chamber.

**Drop-off**

6. A Metal electrode (100/5 nm Au/Ti) was fabricated on the $SiO_2$/Si substrate by standard e-beam lithography, evaporation, and lift-off. The substrate was cleaned by piranha solution for 1 min followed by ultrapure water rinse. The heterostructure on Elvacite2552C + ion was aligned and placed onto the $SiO_2$/Si at 90 °C. Elvacite2552C + ion was melted and the heterostructure was transferred to the $SiO_2$/Si. Finally, the Elvacite2552C + ion was removed by chloroform.

**Laser-based micro-focused angle-resolved photoemission spectroscopy (μ-ARPES) measurement**

The samples are pumped and sealed in an ICF-70 nipple with a gate valve, and transported to the ultrahigh vacuum chamber in about 1 hour. Before μ-ARPES measurement, the samples are annealed around 200 °C for ~10 hours in the ultrahigh vacuum. The laser-based μ-ARPES measurement is performed by using a combination of the hemispherical analyzer (DA30, Scienta Omicron Inc.) and the fourth-harmonic generation with $h\nu$ = 6.42 eV of Ti:sapphire laser radiation (Verdi V-18 and MIRA-HP, Coherent Inc.) obtained by the frequency convertor (HarmoniXX, APE Inc.). The laser incident light is focused by the optical lens system[40] (NTT Advanced Technology corporation) equipped outside with the ultra-high vacuum chamber of the ARPES. The spot size is estimated to be approximately 20 μm at the measurement position by the knife-edge scanning from the Au contact to the $SiO_2$ substrate[11]. On transforming emission angle to the momentum, we adopted the work function of 4.4 eV referring to the ARPES data for bulk $WTe_2$ recorded by higher energy light source[41,42] for 5-layer $WTe_2$ recorded using $h\nu$ = 6.42 eV, where the band dispersions exhibited nearly comparable momentum and band gradient to the data. The total energy resolution was set to 3 meV. During measurement, a sample manipulator temperature was kept below 20 K. The actual temperature of the 2–5 layer $WTe_2$ flakes were evaluated from the width of the Fermi cutoff to be 27(8), 31(5), 23(3), and 25(4) K, respectively. Except Fig. 2i, all ARPES images in this article represent the sums of the ARPES intensities taken with the *s*- and *p*-polarized light.

**Band Calculation**

The first principles band calculations are performed in the frame work with the DFT using the Vienna Ab initio Simulation Package (VASP)[43,44]. To construct the slab-model of 2–5 layer $WTe_2$, the experimental values of the lattice constants and atomic coordinates[33] were used. All calculations were performed within non-spin-polarized approximation with spin–orbit correction (SOC). To approximate the electron exchange-correlation energy, the PBE generalized gradient approximation[32], and the HSE hybrid functional (HSE06)[25] were used. We employed $E_c$ = 300 eV as the cutoff energy for the planewave basis set, and $N_k$ = 12 × 12 × 1 as the number of *k*-points for the self-consistent calculation. The band dispersion with the HSE06 functional was obtained by the Wannier interpolation[45].

A. A., & Yates, J. R. Wannier90 as a community code: new features and applications. J. Phys.: Condens. Matter 32, 165902 (2020).

**Acknowledgement**

We thank Y. Yanase for fruitful discussion. This research was partly supported by a CREST project (Grant Nos. JPMJCR15F3, JPMJCR16F2 and JPMJCR20B4) from the Japan Science and Technology Agency (JST) and Japan Society for the Promotion of Science KAKENHI (Grants-in-Aid for Scientific Research) (Grant Nos. JP20H00127, JP20H00354 and JP20H01834).

**Author contributions**

M.S., S.M., T.M, and K.I. conceived the project. S.O. and T.S. grew the bulk $WTe_2$ crystal. K.W. and T.T. grew the bulk hBN crystal. T.N., R.A. and T.S. performed the band calculations. Y. T., M. S. and S. M. fabricated the few-layer $WTe_2$ samples for the μ-ARPES. M.S., Y.T. and A.O. performed the laser μ-ARPES measurement and analyzed the data. M.S., S.M. and K. I. wrote the manuscript with input from Y.T., T.S., T.N. and T.M. All authors contributed to the scientific discussions.

**Data availability**

Source data are provided with this paper. All other data that support the plots within this paper and other findings of this study are available from the corresponding author upon reasonable request.

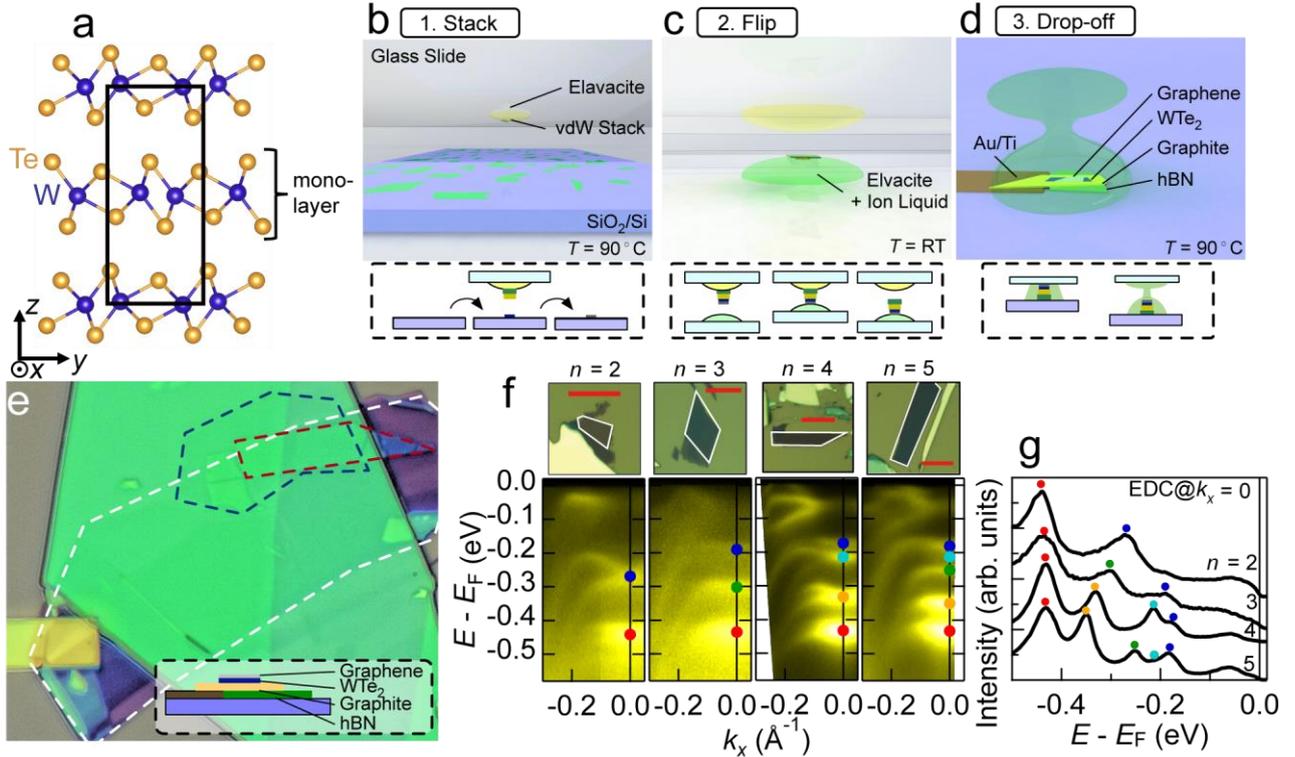

**Figure 1| Fabrication of graphene-encapsulated WTe$_2$ flake samples for ARPES.**
**a,** Crystal structure of WTe$_2$ with the orthogonal (*xyz*) axes. The black rectangle indicates the unit cell of the bulk crystal. **b,** hBN, graphite, WTe$_2$, and graphene flakes were sequentially picked-up by a polymer substrate (Elvacite2552C) at $T$ = 90 °C, **c,** The assembled hBN/graphite/WTe$_2$/graphene heterostructure was transferred to the Elvacite2552C mixed with adhesion additives (ionic liquid) at room temperature. **d,** The heterostructure was dropped onto a SiO$_2$/Si substrate with the prepatterned metal (Au/Ti) electrode at $T$ = 90 °C. **e,** Optical microscope image of the fabricated graphene/WTe$_2$/graphite/hBN heterostructure sample. **f,** Upper panels: The optical microscope images of the 2–5 layer WTe$_2$ flakes (shown with white frames) used for ARPES. The red bar represent the 10 μm length. Lower panels: Obtained ARPES images for 2–5 layer WTe$_2$ recorded along the $k_x$ direction. The makers (●) indicate the positions of the intensity peaks at $k_x$ = 0 (see **g**). **g,** Energy distribution curves (EDCs) at $k_x$ = 0 obtained with an integral width of 0.02 Å$^{-1}$, presented for 2–5 layers. The markers (●) are the positions of the intensity peaks.

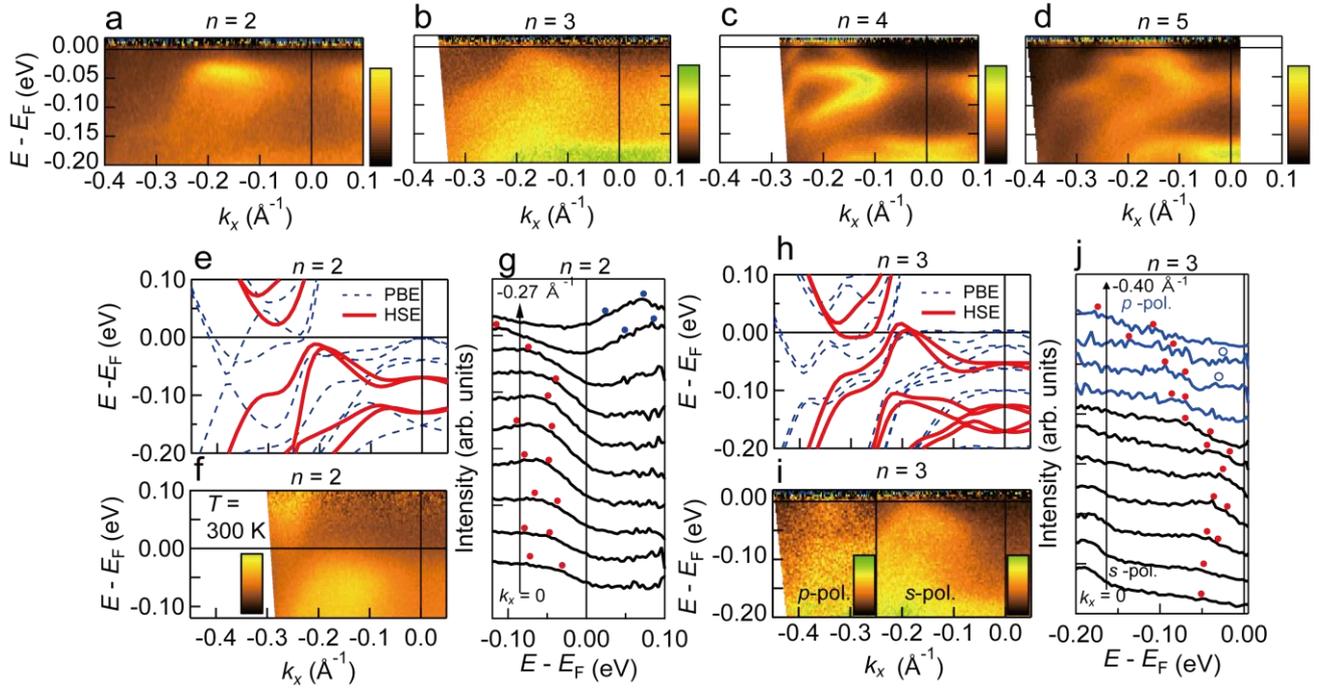

**Figure 2| Evolution of the band dispersions near the Fermi level**

**a-d,** ARPES images near the Fermi level for 2–5 layer WTe$_2$. To remove the effect of Fermi-Dirac cutoff, all ARPES images in panels **a–d**, **f** and **i** are divided by the Fermi–Dirac distribution functions convolved with the Gaussian function with full width at half maximum (FWHM) of 3 meV. **e,** Band calculations on 2-layer WTe$_2$ obtained by using the GGA-PBE (blue dashed curves) and the HSE hybrid functional (red solid curves) as implemented in the Vienna Ab initio Simulation Package (VASP) (see Methods). **f,** ARPES image for 2-layer WTe$_2$ taken at room temperature. **g,** EDCs extracted from the ARPES image for 2-layer WTe$_2$ in **f**. The red and blue markers represent the positions of the intensity peaks for the valence and conduction bands, respectively. **h**, Band calculations for 3-layer WTe$_2$ obtained by using GGA-PBE (blue dashed curves) and the HSE hybrid functional (red solid curves) method. **i,** ARPES image for 3-layer WTe$_2$ presented by combining the data using $p$- ($k_x < -0.25$ Å$^{-1}$) and $s$- ($k_x > -0.25$ Å$^{-1}$) polarized lights. **j,** EDCs extracted from the ARPES image for 3-layer WTe$_2$ in **i**. The red filled (blue empty) circle markers indicate the positions of the intensity peaks for valence (conduction) bands.

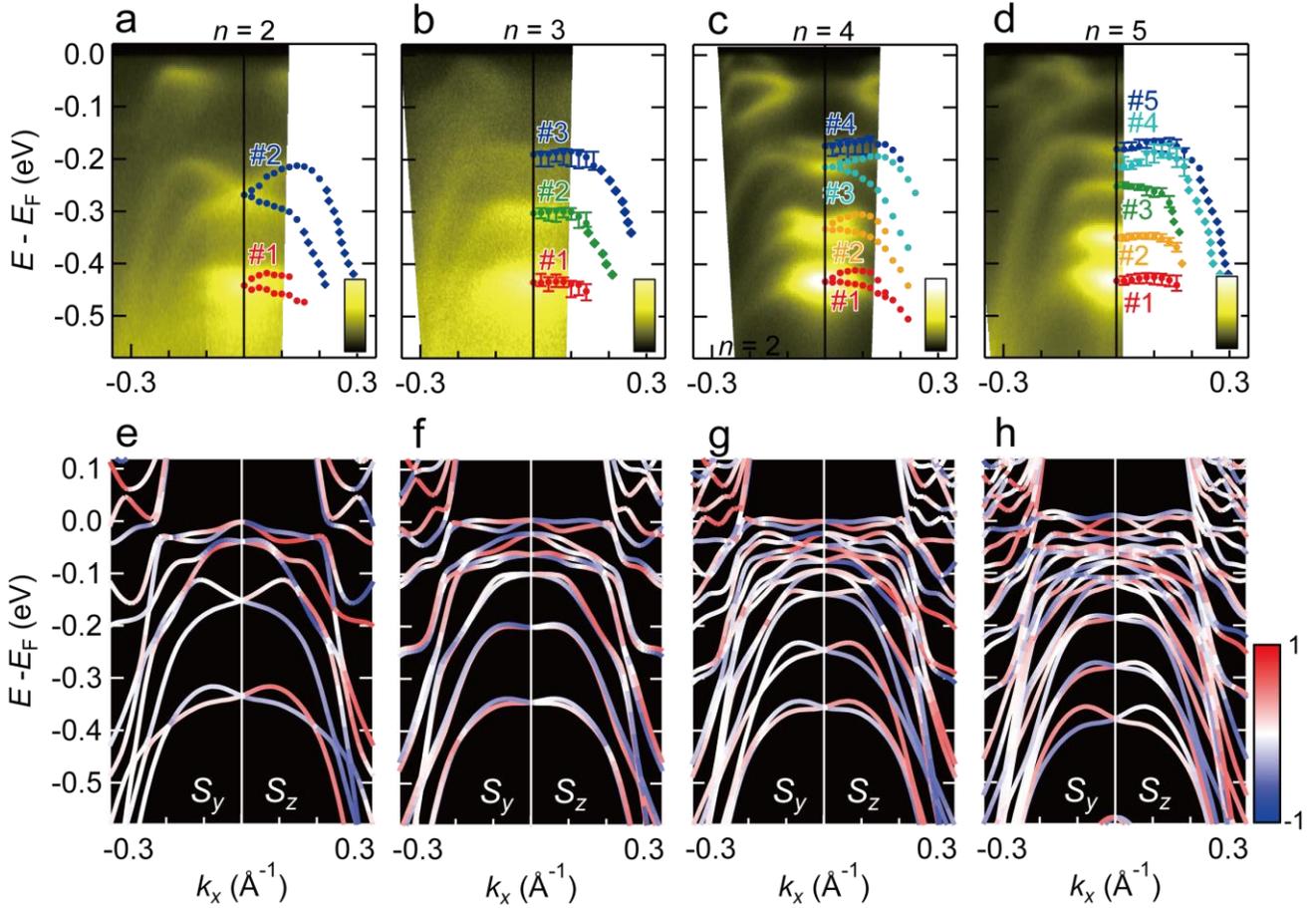

**Figure 3| Layer-number-dependent band dispersions in 2–5 layer WTe$_2$**
**a-d,** ARPES images of 2–5 layer WTe$_2$ recorded along the $k_x$-direction. Symbols ● and ♦ represent the peak positions obtained by fitting the EDCs and the momentum distribution curves (MDCs), respectively. The markers accompanying the bars represent the results considering both single and double component fitting (see text). **e-h,** Band calculations along the $k_x$-direction based on density functional theory (DFT) by using GGA-PBE (see Method). The slab structures are consisted by using the atomic coordinates of the bulk crystal[32] without any relaxations. The band dispersions in $k_x < 0$ and $k_x > 0$ are indicated by the color of the spin y- and z-components ($S_y$ and $S_z$), respectively, according to the color scale on the right of panel **h**. The spin axes are defined by the orthogonal axes (*xyz*) shown in Fig. 4c,d.

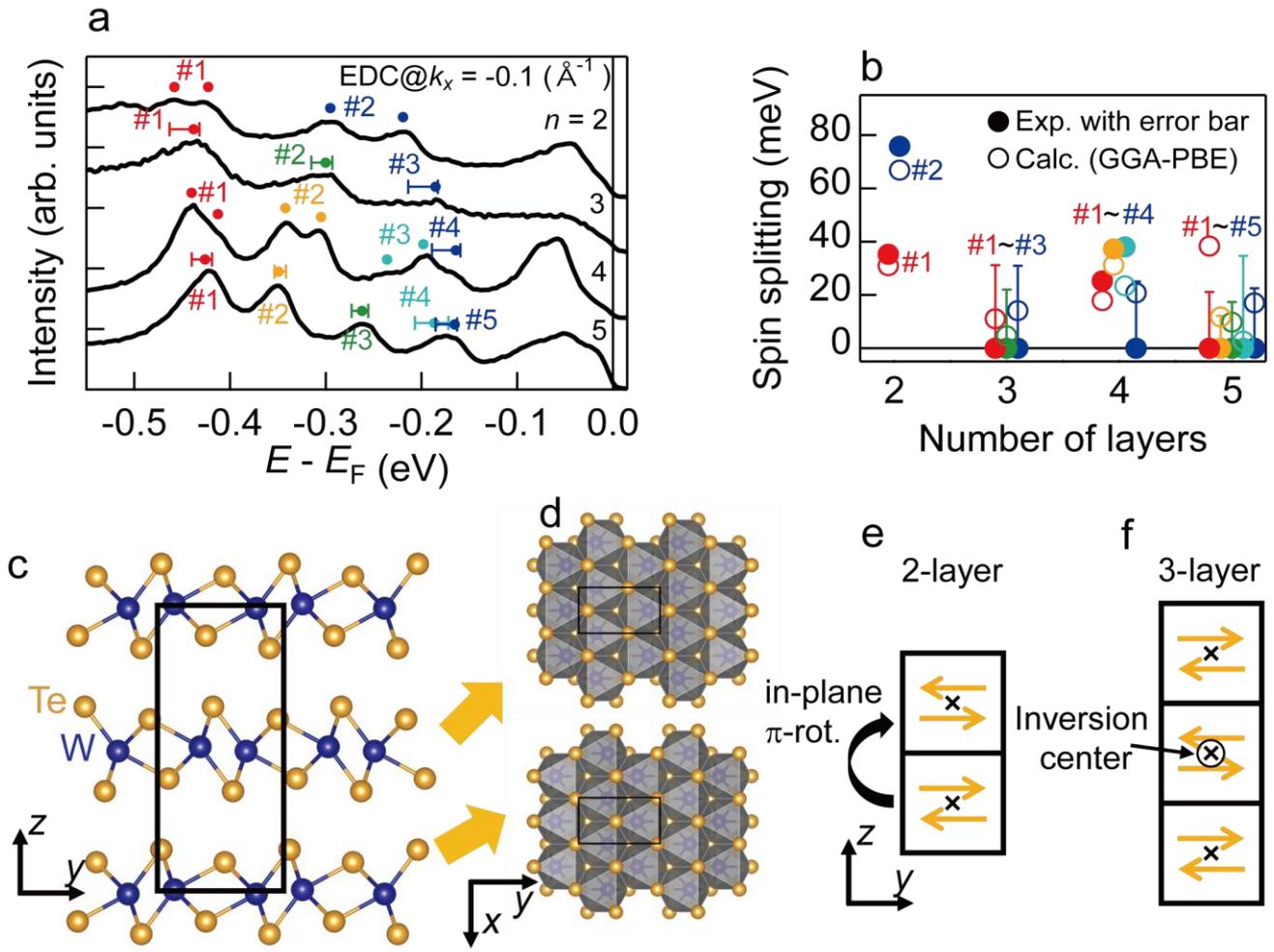

**Figure 4| Origin of the even–odd effect in the layer number dependence**
**a,** EDCs at $k_x = -0.1$ Å$^{-1}$ for 2–5 layer WTe$_2$. **b,** Layer-number-dependence of the energy of the spin band splitting obtained by the ARPES experiment (●) and the DFT calculation (○). **c,** Side view of the bulk crystal structure of WTe$_2$. The black rectangle represents the unit cell. **d,** Top views of adjacent monolayer WTe$_2$ layers. **e,f,** Simplified schematic models of 2- and 3-layer WTe$_2$ where the structural asymmetry within the monolayer is ignored[22] (see text for the detail). The black block and orange arrows represent each unit of WTe$_2$ layer and the directions of Te triangular networks within the layer. The cross markers (×) represent the C$_2$ axis within each WTe$_2$ layer.